\documentclass[10pt,twoside]{article}
\usepackage{ICBStemplate}
\almvol{00}
\almttone{Frontiers of Science Awards}
\almtttwo{for Physics (2025)}
\firstpage{1}
\usepackage[english]{babel}
\usepackage[utf8]{inputenc}
\usepackage{amssymb} %
\usepackage{newunicodechar} %
\newunicodechar{⊙}{\ensuremath{\odot}} %
\usepackage{physics}
\usepackage[dvipsnames, usenames]{xcolor}
\definecolor{linkcolor}{rgb}{0.0,0.3,0.5}
\usepackage[pdfpagelabels=true, unicode, colorlinks=true, linkcolor=linkcolor, citecolor=linkcolor, filecolor=linkcolor,urlcolor=linkcolor, pdfusetitle]{hyperref}
\usepackage[numbers,compress]{natbib}  

\usepackage{etoolbox}

\usepackage{multicol}
\usepackage{tabu}
\usepackage{amsmath}
\usepackage{amssymb}
\usepackage{graphicx}
\usepackage{xcolor}
\usepackage{anyfontsize}
\usepackage{enumitem}

\numberwithin{equation}{section}

\usepackage{titlesec}
\usepackage[font=small]{caption}
\titleformat{\section}
  {\large\bfseries} %
  {\thesection}     %
  {1em}             %
  {}                %

\titleformat{\subsection}
  {\normalsize\bfseries} %
  {\thesubsection}     %
  {0.5em}             %
  {}                %

\makeatletter
\renewenvironment{thebibliography}[1]
     {\section*{\refname}%
      \interlinepenalty=10000 %
      \@mkboth{\MakeUppercase\refname}{\MakeUppercase\refname}%
      \list{\@biblabel{\@arabic\c@enumiv}}%
           {\settowidth\labelwidth{\@biblabel{#1}}%
            \leftmargin\labelwidth
            \advance\leftmargin\labelsep
            \@openbib@code
            \usecounter{enumiv}%
            \let\p@enumiv\@empty
            \renewcommand\theenumiv{\@arabic\c@enumiv}}%
      \sloppy
      \clubpenalty4000
      \@clubpenalty \clubpenalty
      \widowpenalty4000%
      \sfcode`\.=\@m}
     {\def\@noitemerr
       {\@latex@warning{Empty `thebibliography' environment}}%
      \endlist}
\makeatother

\begin{document}

\markboth{\hfill{\rm Gerosa, Foroni, Fumagalli, Berti} \hfill}{\hfill {\rm Coincident morphological transitions in precessing black-hole binaries \hfill}}

\title{\large Coincident morphological transitions \\ in precessing black-hole binaries\\}

\author{\sc \normalsize \vspace{0.1cm} Davide Gerosa, Giulia Foroni, Giulia Fumagalli, Emanuele Berti}

\begin{abstract}
\small 
We present new insights into the phenomenology of post-Newtonian spin precession in black-hole binaries. Using multi-timescale methods, previous work has shown that the precession and nutation dynamics in such systems can be classified into so-called spin morphologies—mutually exclusive regions that partition the configuration space and characterize the motion of the black-hole spins relative to the binary’s angular momentum. Radiation reaction can induce secular transitions between different morphology classes, which are generic occurrences during the inspiral of black-hole binaries. In this contribution, we systematically explore a more restrictive class of solutions in which multiple morphological transitions occur concurrently, i.e., within the same precession cycle. We find that all such cases can be mapped and characterized analytically, and we confirm these findings through numerical integrations. These coincident transitions correspond to extreme spin configurations in black-hole binaries with potential observational signatures in gravitational-wave astronomy.
\end{abstract}

\maketitle

\section{Introduction}

Black-hole (BH) binary spin precession is a distinctive feature of the relativistic two-body problem. As two BHs orbit each other and inspiral due to gravitational-wave (GW) radiation reaction, spin-orbit and spin-spin couplings cause the binary orbital angular momentum and the two BH spins to precess around the direction of the total angular momentum of the system \cite{1994PhRvD..49.6274A}. In GW astronomy, the detection of spin precession provides a valuable tool to constrain the astrophysical environments in which BHs form and evolve ---both in the stellar-mass regime targeted by LIGO/Virgo/KAGRA \cite{2025arXiv250818083T} and in the supermassive regime targeted by LISA \cite{2024arXiv240207571C}. Spin signatures are related to specific astrophysical details that set the formation and evolution of BH binaries, most notably mass transfer events, core-evelope interactions in massive stars, tidal interactions, disk accretion \cite{2022PhR...955....1M,2021hgwa.bookE..16M,2021FrASS...8....7S,2023LRR....26....2A}, as well as the occurrence of repeated mergers \cite{2017PhRvD..95l4046G,2017ApJ...840L..24F,2021NatAs...5..749G}.

The study of precessing BH binaries, at least in the post-Newtonian (PN) regime, is greatly simplified by two timescale separations. The orbital timescale $t_{\rm orb}$ is much shorter than the spin-precession timescale $t_{\rm pre}$, which is itself much shorter than the radiation-reaction timescale $t_{\rm rad}$. This hierarchy allows one to average the evolutionary equations over both the orbital \cite{1964PhRv..136.1224P} and precession timescales \cite{2015PhRvD..92f4016G}, and to introduce the inspiral in a quasi-adiabatic fashion.

In this framework, the evolution of the spins on the precession timescale $t_{\rm pre}$ can be treated analytically up to 2PN order \cite{2021arXiv210610291K,2023PhRvD.108b4042G}. In recent years, this has enabled both the authors and other researchers to ``maximize'' spin-precession effects—that is, to explore the parameter space for peculiar configurations in which relativistic spin couplings most strongly influence the binary dynamics. In some cases, previously discovered effects have been reinterpreted and better understood using this multiple-timescale approach. These findings include the classification of BH binaries into “spin morphologies” \cite{2015PhRvD..92f4016G} (reviewed below), as well as the identification of spin-orbit resonances \cite{2004PhRvD..70l4020S,2010PhRvD..81h4054K} or Cassini states \cite{2016MNRAS.457L..49C}, dynamical instabilities \cite{2015PhRvL.115n1102G,2016PhRvD..93l4074L,2020PhRvD.101l4037M}, nutational resonances \cite{1994PhRvD..49.6274A,2017PhRvD..96b4007Z}, and extreme nutations \cite{2019CQGra..36j5003G,2016PhRvD..93d4031L} (also reviewed below). Multi-timescale solutions of spin-precession dynamics are now incorporated into some state-of-the-art waveform models \cite{2021arXiv210610291K,2019PhRvD.100b4059K,2020PhRvD.102f4001P} used for GW parameter estimation.

In this paper, we continue along this line of research, identifying specific regions of the parameter space where precession effects are particularly prominent. In particular, we show that the PN dynamics admits special configurations in which both BH spins become instantaneously aligned with the orbital angular momentum during a single precession cycle. Crucially, these are not binaries with permanently aligned spins: they are fully precessing systems, where both spins oscillate so extensively that they transiently reach aligned configurations. We demonstrate that these configurations are generalizations of the morphological transitions first highlighted in Refs.~\cite{2015PhRvD..92f4016G,2015PhRvL.114h1103K}, where two coincident transitions occur at the same orbital separation during the binary inspiral. We show that these peculiar configurations can be identified and characterized entirely analytically, and we confirm these findings with targeted numerical integrations of the PN equations of motion.

\section{Morphological transitions}

In this paper, we use spin-precession equations accurate up to 2PN~\cite{2008PhRvD..78d4021R} together with precession-averaged radiation-reaction equations accurate up to 1PN~\cite{2015PhRvL.114h1103K}. A recent reformulation of the problem, which we also adopt here, was presented in Ref.~\cite{2023PhRvD.108b4042G} and is implemented in the {\sc precession} module for the Python programming language~\cite{repo} (see Ref.~\cite{2022PhRvD.106b3001J} for another public implementation of the same formalism). We restrict our study to BHs on quasi-circular orbits, which are well motivated astrophysically (but see Refs.~\cite{2020PhRvD.102l3009Y,2023PhRvD.108l4055F,2025arXiv250507238S} for generalizations to BHs on eccentric orbits and Ref.~\cite{2023PhRvD.108d3018L} for a generalization to neutron-star binaries).

Let us consider a BH binary with component masses $m_{1,2}$, mass ratio $q=m_2/m_1\leq 1$, total mass $M=m_1+m_2$, dimensionless spin magnitudes $\chi_{1,2}$, and  separation $r$.  The mutual direction of the spins and the orbital angular momentum can be described by the polar angles $\theta_{1,2}$ between the spins and the orbital angular momentum and the azimuthal angle $\Delta\Phi$ between the projections of the two spins onto the orbital plane. These three angles $\theta_{1}$, $\theta_{2},$ and $\Delta\Phi$ all vary on the precessional timescale. More conveniently, one can use a parametrization that respects the timescale separation of the problem. In particular, we use
\begin{align}
\chi_{\rm eff}&= \frac{\chi_1\cos\theta_1 + q \chi_2 \cos\theta_2}{1+q}\,,
\label{chieff}\\
 \kappa  &= \frac{1}{(1+q)^2}\bigg\{{\chi_1 \cos\theta_1  + q^2  \chi_2\cos\theta_2 }{} 
 \notag \\ 
 &+ \frac{1}{2 q \sqrt{r/M}}\Big[ \chi_1^2 
+ 2 q^2 \chi_1 \chi_2  (\cos\theta_1 \cos\theta_2 + \cos\Delta\Phi \sin\theta_1
   \sin\theta_2) +q^4 \chi_2^2 \Big]\bigg\},
 \label{kappa}\\
 \delta\chi &= \frac{\chi_1\cos\theta_1 - q \chi_2 \cos\theta_2}{1+q}\,,
 \label{deltachi}
\end{align}
where the effective spin $\chi_{\rm eff}$  is a constant of motion~\cite{2008PhRvD..78d4021R}, the asymptotic angular momentum $\kappa$ is constant on the precession timescale and varies only on the radiation-reaction timescale~\cite{2015PhRvD..92f4016G}, and the weighted spin difference $\delta\chi$ encodes the precession dynamics~\cite{2021arXiv210610291K}.

With this parametrization in hand, spin precession can be treated quasi-adiabatically, first solving for the dynamics on the precessional timescale (which is an analytically tractable one-dimensional problem in $\delta\chi$), and then introducing secular variations to capture the inspiral (which reduces to solving an ordinary differential equation for $\dd \kappa / \dd r$). In particular, the weighted spin difference $\delta\chi$ undergoes quasi-periodic oscillations between two endpoints $\delta\chi_{\pm}$. In turn, these endpoints depend on $\kappa$ and $r$, and thus vary slowly as the binary inspirals.
In the following, subscripts $\pm$ indicate\footnote{Some of our previous papers used the total spin $S$ to parametrize the spin-precession dynamics instead of $\delta\chi$. We now prefer using $\delta\chi$ because it allows for taking regular limits when $q\to 1$ \cite{2023PhRvD.108b4042G,2017CQGra..34f4004G}. Because $\dd\delta\chi/\dd S\leq 0$, one has that $\delta\chi_\pm$ corresponds to $S_\mp$. This should be taken into account when comparing the notation used here to that of previous works.}  quantities evaluated at $\delta\chi_\pm$.

The dynamics of BH binaries on the precessional timescale can be classified into discrete classes according to the value of $\Delta\Phi$ at the endpoints $\delta\chi_{\pm}$. These were first identified in Refs.~\cite{2015PhRvD..92f4016G,2015PhRvL.114h1103K} and dubbed ``spin morphologies'' (but see Ref.~\cite{2004PhRvD..70l4020S} for earlier hints). One always has that $\sin\Delta\Phi_\pm=0$, leading to four distinct classes~\cite{2023PhRvD.108b4042G}:
\begin{itemize}
\item[$\bullet$]~~L0: $\Delta\Phi_-=\Delta\Phi_+=0$,
\item[$\bullet$]~~L$\pi$: $\Delta\Phi_-=\Delta\Phi_+=\pi$,
\item[$\bullet$]~~C$+$: $\Delta\Phi_-=0$ and $\Delta\Phi_+=\pi$,
\item[$\bullet$]~~C$-$: $\Delta\Phi_-=\pi$ and $\Delta\Phi_+=0$,
\end{itemize}
where ``C'' stands for circulation and ``L'' stands for libration.
Previous studies, including our own Refs.~\cite{2015PhRvD..92f4016G,2019CQGra..36j5003G,2015PhRvL.114h1103K}, generically refer to both C$+$ and C$-$ as a single C morphology. It was later realized \cite{2023PhRvD.108b4042G} that the two configurations are indeed different, and such a difference turns out to be crucial for the present study. 

\begin{figure}[t]\centering
\includegraphics[width =0.6\columnwidth]{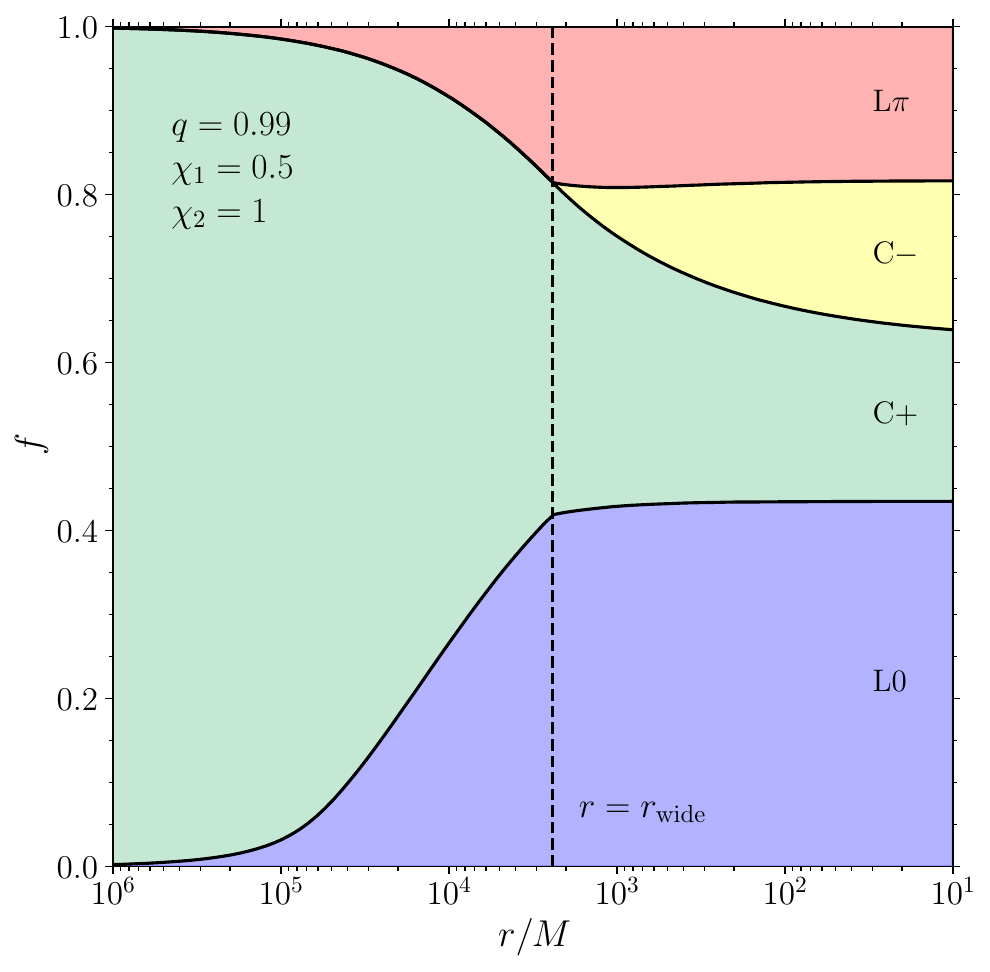}
\caption{Fraction $f$ of BH binaries in each of the four spin morphologies: L$0$ (blue), L$\pi$ (red), C$+$ (green), and C$-$ (yellow). We consider a set of $10^6$ sources with $q=0.99$, $\chi_1=0.5$, $\chi_1=1$, and isotropic spin orientations. The vertical dashed line indicates the critical $r=r_{\rm wide}$ from Eq.~(\ref{rwide}) where the C$-$ morphology becomes accessible.}
\label{distrfour}
\end{figure}

The spin morphologies characterize the precession dynamics on $t_{\rm pre}$, but binaries can (and generically will) transition between different morphologies as they inspiral toward merger on $t_{\rm rad}$~\cite{2015PhRvD..92f4016G,2015PhRvL.114h1103K}.
As an example, Fig.~\ref{distrfour} shows the morphology distribution for a set of BH binaries with $q = 0.99$, $\chi_1 = 0.5$, and $\chi_2 = 1$. These binaries are initialized with isotropic spin directions at $r = \infty$ and evolved down to $r = 10M$, where the PN approximation is likely to break down~\cite{2008PhRvD..77l4006Y,2011PhRvD..84b4029Z}. All binaries belong to the C$+$ morphology at large separation and may transition toward either L0 or L$\pi$. In rarer cases, a second transition can occur before merger, with the binary evolving into C$-$.  
While the first transitions from circulation to libration are generic features of the BH binary problem, the second transitions ---from libration back to circulation--- occurs only for binaries with nearly equal masses and unequal spin magnitudes \cite{2015PhRvD..92f4016G}.
For the case shown in Fig.~\ref{distrfour}, these second transitions start occurring at $r \simeq 2400M$. As shown in this paper, this key location in the inspiral of BH binaries can actually be predicted analytically.

Morphological transitions corresponds to discontinuities of $\Delta\Phi$ at one of the endpoints $\delta\chi_\pm$. Much like the longitude is ill-defined at the Earth's poles, the azimuthal angle $\Delta\Phi$ is ill-defined whenever one of the two spins is co- or counter-aligned with the binary orbital angular momentum. In principle, there are eight such conditions, namely when either $\cos\theta_{1\pm}$ or $\cos\theta_{2\pm}$ is equal to $\pm 1$. However, one has that \cite{2015PhRvD..92f4016G,2023PhRvD.108b4042G}
\begin{align}
\frac{\dd \cos\theta_1}{\dd \delta\chi}\geq0\,,
\qquad
\frac{\dd \cos\theta_2}{\dd \delta\chi}\leq0\,,
\end{align}
which implies only four of these eight conditions are viable. These are $\cos\theta_{1-}= -1$, $\cos\theta_{1+}=+1$, $\cos\theta_{2-}=+1$, and $\cos\theta_{2+}=-1$ and mark the occurrence of a morphological transition.

\section{Coincident transitions: six cases}

The phenomenology of the spin morphologies and their related transitions has been studied extensively in  Ref.~\cite{2015PhRvD..92f4016G}, to which we refer for further details. %
In this work, we specifically look for cases where there are two coincident transitions, which turns out can all be solved analytically. While hints of such occurrences are present in, e.g., Refs.~\cite{2015PhRvL.115n1102G,2019CQGra..36j5003G}, a systematic analysis like the one carried out here has not been presented elsewhere and represents a novel contribution.

From the four conditions above, one can construct six possibilities where concurrent transitions could, in principle, take place:
\begin{enumerate}[label=\bf \arabic{enumi}.]
\item $\;\;\cos\theta_{1-}\!=\!-1, \cos\theta_{1+}\!=\!+1$,\label{wideprimary}
\item $\;\;\cos\theta_{2-}\!=\! +1, \cos\theta_{2+}\!=\! -1$, \label{widesecondary}
\item $\;\;\cos\theta_{1-}\!=\!-1, \cos\theta_{2+}\!=\! -1$,\label{coincidentprimarysmall}
\item $\;\;\cos\theta_{1+}\!=\!+1, \cos\theta_{2-}\!=\! +1$, \label{coincidentsecondarysmall}
\item $\;\;\cos\theta_{1+}\!=\!+1, \cos\theta_{2+}\!=\! -1$, \label{updown}
\item $\;\;\cos\theta_{1-}\!=\!-1, \cos\theta_{2-}\!=\! +1$. \label{impossible}
\end{enumerate}

Neither $\chi_{\rm eff}$ nor $\kappa$ vary on the precession timescale and,  therefore, their values must be the same when evaluated at either $\delta\chi_-$ or $\delta\chi_+$. From Eqs.~(\ref{chieff}) and (\ref{kappa}), one must impose that:
\begin{align}
&\chi_1\cos\theta_{1-} + q \chi_2 \cos\theta_{2-} = \chi_1\cos\theta_{1+} + q \chi_2 \cos\theta_{2+}
\label{condition1}
\end{align}
and
\begin{align}
&q \chi_1 \chi_2 (\cos\theta_{1-} \cos\theta_{2-} + \cos\Delta\Phi_+ \sin\theta_{1-} \sin\theta_{2-})  
\notag\\  
 &\qquad
 +\sqrt{r/M} ( \chi_1 \cos\theta_{1-}  + q^2 \chi_2\cos\theta_{2-} )
 \notag \\ 
  &=q \chi_1 \chi_2 (\cos\theta_{1+} \cos\theta_{2+} + \cos\Delta\Phi_+ \sin\theta_{1+} \sin\theta_{2+})  
\notag\\ 
& \qquad
+\sqrt{r/M} ( \chi_1 \cos\theta_{1+}  + q^2 \chi_2\cos\theta_{2+} )\,.
\label{condition2}
\end{align}

\begin{figure*}[p!]
$\!\!$
\includegraphics[height=0.53\textwidth,page=1]{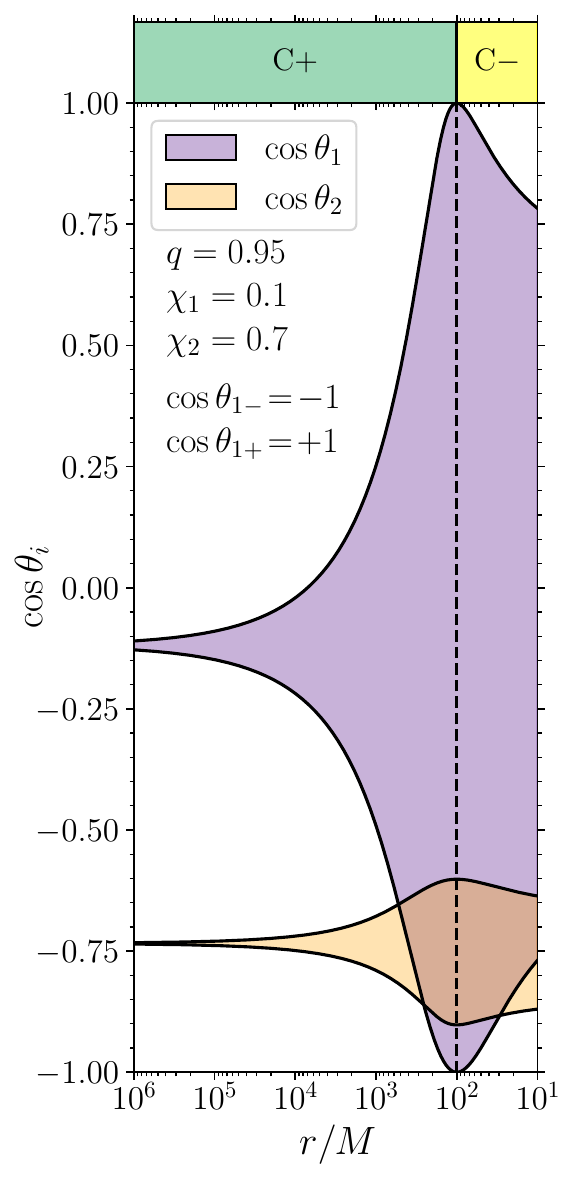}
$\!\!\!\!\!$
\includegraphics[height=0.53\textwidth,page=2]{doubletransitions.pdf}
$\!\!\!\!\!$
\includegraphics[height=0.53\textwidth,page=3]{doubletransitions.pdf}
$\!\!\!\!\!$
\includegraphics[height=0.53\textwidth,page=4]{doubletransitions.pdf}$\!\!$
\\
$\!\!$
\includegraphics[height=0.53\textwidth,page=5]{doubletransitions.pdf}
$\!\!\!\!\!$
\includegraphics[height=0.53\textwidth,page=6]{doubletransitions.pdf}
$\!\!\!\!\!$
\includegraphics[height=0.53\textwidth,page=7]{doubletransitions.pdf}
$\!\!\!\!\!$
\includegraphics[height=0.53\textwidth,page=8]{doubletransitions.pdf}$\!\!$
\caption{
Precession-averaged evolution of the tilt angles $\theta_1$ (purple) and $\theta_2$ (orange) from an initial separation of $r = 10^6 M$ down to $r = 10 M$. These angles oscillate within the envelopes shown in the figure, which correspond to the binary configurations at the endpoints $\delta\chi_\pm$. The colored bar at the top of each panel indicates the spin morphology of the system: L$0$ (blue), L$\pi$ (red), C$+$ (green), and C$-$ (yellow). A change in morphology (dashed lines) occurs whenever one of the two spins becomes aligned or anti-aligned with the orbital angular momentum, i.e., when $\cos\theta_{1,2} = \pm 1$.
The various panels correspond to different values of $q$, $\chi_1$, $\chi_2$, $\chi_{\rm eff}$, and $\kappa$, chosen to illustrate all cases of double and triple transitions analyzed in this paper. In particular, from left to right and top to bottom, the panels show: wide nutation of the primary BH (case~\ref{wideprimary}); wide nutation of the secondary BH (case~\ref{widesecondary}); coincident transitions with small primary spin (case~\ref{coincidentprimarysmall}); coincident transitions with small secondary spin (case~\ref{coincidentsecondarysmall}); up-down binaries transitioning from circulation to libration (case~\ref{updown}); up-down binaries transitioning from libration to circulation (case~\ref{updown}); triple transition with small primary spin; and triple transition with small secondary spin.
}
\label{doubletr}
\end{figure*}

We now plug each of the six conditions above into Eqs.~(\ref{condition1}) and (\ref{condition2}). Some examples are presented in Fig.~\ref{doubletr}, where we show the secular evolution of the tilt angles $\theta_{1}$ and $\theta_{2}$ under radiation reaction. This figure closely parallels Fig. 11 in Ref.~\cite{2015PhRvD..92f4016G}. Much like that figure provides a comprehensive overview of all cases where zero or one transition can occur, Fig.~\ref{doubletr} in this paper maps out all possible cases involving double and triple transitions. Together, they complete the exploration of morphological transitions in spinning BH binaries.

\subsection{$\boldsymbol{\cos\theta_{1-}\!=\!-1, \cos\theta_{1+}\!=\!+1}$: Wide nutation}
The conditions  $\cos\theta_{1-}=-1, \cos\theta_{1+}=+1$ correspond to the primary BH spin oscillating from full alignment to full anti-alignment in a single precession cycle. This is the ``wide nutation'' phenomenon we first described in Ref.~\cite{2019CQGra..36j5003G}. 
Solutions can be found only if~$\!$\footnote{
In this paper, we often make use of the fact that the following two sets of inequalities have the same solutions:
\begin{align}\notag
\begin{cases}
r_{\rm wide}\geq M\,,\\
\chi_1 \lessgtr \chi_2\,,
\end{cases}
\quad \Longleftrightarrow\qquad
\begin{cases}
r_{\rm wide}\geq M\,,\\
\chi_1 \lessgtr q \chi_2\,.
\end{cases}
\end{align}
While the right set is most stringent, we use the left set for simplicity, as was also done in Ref.~\cite{2019CQGra..36j5003G}.}
\begin{align}
\left\{
\begin{array}{l@{\hspace{2pt}}l}
r&\leq r_{\rm wide}\,,\\
\chi_1&\leq \chi_2\,,
\\
\displaystyle
\chi_{\rm eff} &=  \displaystyle -\frac{1-q}{1+q} \sqrt{r/M}\,,
\\ \displaystyle
\kappa &= \displaystyle -\frac{q(1-q) 
   \sqrt{r/M}}{(1+q)^2} +  \frac{(1-2 q) \chi_1^2 + q^4 \chi_2^2}{2 q (1+q)^2 \sqrt{r/M}}\,,
\end{array}
\right.
\end{align}
where
\begin{equation}
r_{\rm wide} = \left(\frac{\chi_1 - q\chi_2}{1-q}\right)^2 M\,.
\label{rwide}
\end{equation} 

Wide nutation is indeed a case of two coincident morphological transitions, one occurring at $\delta\chi_-$ and one occurring at $\delta\chi_+$. A binary initially in the C$+$ will transition directly to the C$-$ morphology, without an intermediate phase of libration.\footnote{In principle, two transitions occurring at $\delta\chi _-$ and $\delta\chi_+$ could also result in the spin morphology evolving between the two librating classes without an intermediate phase of circulation. While we were not able to exclude this possibility analytically, we could not identify any of these cases even after extensive numerical investigations.}

Figure~\ref{doubletr} shows an example with $q=0.95$, $\chi_1=0.1$, $\chi_2=0.7$ where wide nutation happens at $r=100M$ for $\chi_{\rm eff}\simeq = -0.26$ and $\kappa\simeq-0.12$; for these parameters $r_{\rm wide}\simeq 128 M$.

\subsection{$\boldsymbol{\cos\theta_{2-}\!=\! +1, \cos\theta_{2+}\!=\! -1}$: Wide nutation}

This second case where $\cos\theta_{2-}\!=\! +1, \cos\theta_{2+}\!=\! -1$ leads to wide nutation for the secondary BH~\cite{2019CQGra..36j5003G}. Solutions exist only if 
\begin{align}
\left\{
\begin{array}{l@{\hspace{2pt}}l}
r&\leq r_{\rm wide}\,,\\
\chi_2&\leq \chi_1\,,\\
\chi_{\rm eff} &=\displaystyle  \frac{1-q}{1+q} \sqrt{r/M}\,,
\\
\kappa &=\displaystyle  \frac{(1-q) 
   \sqrt{r/M}}{(1+q)^2} +  \frac{\chi_1^2 - (2-q) q^3 \chi_2^2}{2 q (1+q)^2 \sqrt{r/M}}\,.
\end{array}
\right.
\end{align}
As above, two transitions happen concurrently, with the morphology evolving from C$+$ to C$-$. 

Figure~\ref{doubletr} shows an example with $q=0.95$, $\chi_1=1$, $\chi_2=0.2$ where wide nutation happens at $r=200M$ for $\chi_{\rm eff}\simeq = 0.36$ and $\kappa\simeq 0.20$; for these parameters $r_{\rm wide}\simeq 262 M$.

\subsection{$\boldsymbol{\cos\theta_{1-}\!=\!-1, \cos\theta_{2+}\!=\! -1}$: Coincident transitions}

The condition $\cos\theta_{1-}\!=\!-1, \cos\theta_{2+}\!=\! -1$ indicates that, as they precess, the two BH spins both become fully anti-aligned with the orbital angular momentum, though at two different points during the same precession cycle.
Equations~(\ref{condition1}) and (\ref{condition2}) admit a trivial solution where $\cos\theta_{1\pm}=\cos\theta_{2\pm}=-1$, which describes a non-precessing system. Less trivially, we find an additional one-parameter family of solutions for
\begin{align}
\left\{
\begin{array}{l@{\hspace{2pt}}l}
r &=  r_{\rm wide}\,,\\
\chi_1 &\leq \chi_2\,,\\
\chi_{\rm eff} &= \displaystyle \frac{\lambda \chi_1- q\chi_2}{1+q}\,,
\\
\kappa &= \displaystyle - 
\frac{\left(\chi_1-q^2 \chi_2\right) \left[(1-q-2q\lambda) \chi_1 +q^2 
  (1+q) \chi_2 \right]}{2 q (1+q)^2 (\chi_1-q \chi_2)}\,,
\end{array}
\right.
\label{coincidentpr}
\end{align}
where $-1\leq\lambda\leq 1$.
 Much like in the wide-nutation cases, here we also have two concurrent transitions taking place at $\delta\chi_-$ and $\delta\chi_+$, which thus corresponds to a direct change in morphology from C$+$ to C$-$. However, while in wide nutation a single BH is responsible for both transitions, here the two BHs are responsible for one transition each. 
 
Figure~\ref{doubletr} shows an example for $q=0.95$, $\chi_1=0.4$, $\chi_2=0.8$, and $\lambda=0.3$. The double transitions happen at $r=r_{\rm wide}\simeq 52 M$ for binaries with $\chi_{\rm eff}\simeq-0.33$ and $\kappa\simeq-0.15$.

\subsection{$\boldsymbol{\cos\theta_{1+}\!=\!+1, \cos\theta_{2-}\!=\! +1}$: Coincident transitions}

This case mirrors the previous one; the only difference is that both BHs evolve through full co-alignment, rather than counter-alignment, with the orbital angular momentum. Besides the trivial solution with $\cos\theta_{1\pm}=\cos\theta_{2\pm}=+1$, we find that the conditions $\cos\theta_{1-}\!=\!+1, \cos\theta_{2+}\!=\! +1$ are satisfied when
\begin{align}
\left\{
\begin{array}{l@{\hspace{2pt}}l}
r &=  r_{\rm wide}\,,\\
\chi_2 & \leq \chi_1\,,\\
\chi_{\rm eff} &= \displaystyle \frac{\chi_1 + q\lambda\chi_2}{1+q}\,,
\\
\kappa &= \displaystyle
\frac{\left(\chi_1-q^2 \chi_2\right) \left[(1+q) \chi_1-q^2 (
   1-q-2 \lambda) \chi_2 \right]}{2 q (1+q)^2 (\chi_1-q \chi_2)}\,,
\end{array}
\right. 
\end{align}
where $-1\leq\lambda\leq 1$.  Once more, the spin morphology transitions from C$+$ to C$-$.

Figure~\ref{doubletr} shows an example for $q=0.95$, $\chi_1=1$, $\chi_2=0.5$, and $\lambda=-0.2$. The double transitions happen at $r=r_{\rm wide}\simeq 110 M$ for binaries with $\chi_{\rm eff}\simeq 0.46$ and $\kappa\simeq 0.25$.

\subsection{$\boldsymbol{\cos\theta_{1+}\!=\!+1, \cos\theta_{2+}\!=\! -1}$: Up-down instability}

This condition is qualitatively different from all previous cases, as both transitions happen at the same endpoint, in this case $\delta\chi_+$. In particular, the condition $\cos\theta_{1+}\!=\!+1, \cos\theta_{2+}\!=\! -1$ corresponds to the so-called ``up-down'' configuration, where the primary (secondary) BH is co-aligned  (counter-aligned) with the orbital angular momentum. Once more, we find the trivial solution $\cos\theta_{1\pm}=- \cos\theta_{2\pm}=1$. An additional precessing solution is present whenever

\begin{align}
\left\{
\begin{array}{l@{\hspace{2pt}}l}
r_{\rm UD-} &\leq r \leq r_{\rm UD+}\,,\\
\chi_{\rm eff} &= \displaystyle\frac{\chi_1- q\chi_2}{1+q}\,,
\\
\kappa &=\displaystyle \frac{\chi_1 - 
  q^2 \chi_2}{(1 + q)^2} \left(1 + \frac{\chi_1 - q^2 \chi_2}{2 q  \sqrt{r/M}} \right)\,.
\end{array}
\right. 
\label{updownsolution}
\end{align}
where
\begin{equation}
r_{\rm UD\pm} = \frac{(\sqrt{\chi_1} \pm \sqrt{q\chi_2})^4}{(1-q)^2} M\,.
\end{equation}
Let us note that $r_{\rm wide} = \sqrt{r_{\rm UD+}r_{\rm UD-}}$ hence $r_{\rm UD-}\leq r_{\rm wide}< r_{\rm UD-}$.
To derive Eq.~(\ref{updownsolution}) we have used $\sin \Delta\Phi_- = 0$, which is generically true \cite{2015PhRvD..92f4016G,2023PhRvD.108b4042G}. In particular, for $r>r_{\rm wide}$ the valid solution is that with $\cos\Delta\Phi_- = +1$ while for $r<r_{\rm wide}$ one has  $\cos\Delta\Phi_- = -1$.

These findings are in full agreement with those of Ref.~\cite{2015PhRvL.115n1102G}, where we first pointed out that up-down binaries are unstable to spin precession whenever $r_{\rm UD-} \leq r \leq r_{\rm UD+}$.
Binaries subject to the coincident transitions investigated here are sources for which the up–down configuration is not stable, but rather a single point in the evolutionary cycle of a generically precessing solution.
Note that, because both conditions are satisfied at the same point $\delta\chi_+$, only the value of $\Delta\Phi_+$ becomes ill-defined, while $\Delta\Phi_-$ remains regular. Morphological transitions involving the up-down configurations therefore reduce to a standard, single transition of the kind explored in detail in Ref.~\cite{2015PhRvD..92f4016G}. In particular, if $\cos\Delta\Phi_- = +1$ (which is only possible at separations $r>r_{\rm wide}$), one has a C$+\to$ L$0$ transition, from circulation to libration. If instead $\cos\Delta\Phi_- = -1$ (which is only possible at separations $r<r_{\rm wide}$), one has a L$\pi\to$ C$-$ transition, from libration to circulation. The calculation strategy presented here, namely that of imposing two coincident transitions, provides yet another proof that up-down binaries are unstable under spin precession (see Ref.~\cite{2015PhRvL.115n1102G} for a complementary argument using effective potentials,  Ref.~\cite{2020PhRvD.101l4037M} for a more direct approach based on perturbation theory, and Ref.~\cite{2021PhRvD.103f4003V} for targeted numerical-relativity simulations, and Ref.~\cite{2023PhRvD.108b4024D} for injections into GW data).

Figure~\ref{doubletr} shows two examples of unstable up-down binaries, both with $q=0.95$, $\chi_1=0.4$, $\chi_2=1$, and $\chi_{\rm eff}=-0.28$: a C$+\to$ L$0$ transition occurring at $r=1000 M$ for $\kappa\simeq-0.131$, and a L$\pi\to$ C$-$ trasition occurring at $r=40M$ for  $\kappa\simeq-0.127$.

\subsection{$\boldsymbol{\cos\theta_{1-}\!=\!-1, \cos\theta_{2-}\!=\! +1}$: Not allowed}

This final case presents two transitions at $\delta\chi_-$. The condition $\cos\theta_{1-}\!=\!-1, \cos\theta_{2-}\!=\! +1$ corresponds to ``down-up'' binaries, where the primary (secondary) BH is counter-aligned (co-aligned). In this case, Eqs.~(\ref{condition1}) and (\ref{condition2}) only admit the non-precession solution with $\cos\theta_{1}=- \cos\theta_{2}=1$. Indeed, the down-up configuration was shown to be stable to spin precession at all separations \cite{2015PhRvL.115n1102G,2020PhRvD.101l4037M}. 

\subsection{The only triple transitions} 

From the conditions above, it is straightforward to identify the even stricter conditions where three morphological transitions coexist. These are very fine-tuned cases, and there are only two of such binaries.
The first is 
\begin{align}
\left\{
\begin{array}{l@{\hspace{2pt}}l}
r =  r_{\rm wide}\,,\\
\chi_1\leq \chi_2\,,\\ \displaystyle
\chi_{\rm eff} = \frac{\chi_1- q\chi_2}{1+q}\,,
\\ \displaystyle
\kappa = -\frac{(\chi_1-q^2 \chi_2) [(1-3q)\chi_1+q^2 (1 + q)\chi_2]}{2 q (1+q)^2 (\chi_1-q\chi_2)}
\end{array}
\right. 
\label{triple1}
\end{align}
and satisfies $\cos\theta_{1-} =-1$, $\cos\theta_{1+}=+1$ and $\cos\theta_{2+}=-1$. These conditions encompass cases \ref{wideprimary}, \ref{coincidentprimarysmall}, and \ref{updown} above, indicating a binary where the primary spin undergoes wide nutation, the morphology transitions from C$+$ to C$-$, and the up-down configuration is part of the precession cycle.
The other configuration is
\begin{align}
\left\{
\begin{array}{l@{\hspace{2pt}}l}
r =  r_{\rm wide}\,,\\
\chi_2\leq \chi_1\,,\\ \displaystyle
\chi_{\rm eff} = \frac{\chi_1- q\chi_2}{1+q}\,,
\\ \displaystyle
\kappa = \frac{(\chi_1-q^2 \chi_2) [(1+q)\chi_1- q^2 (3 - q)\chi_2]}{2 q (1+q)^2 (\chi_1-q\chi_2)}\,.\end{array}
\right. 
\label{triple2}
\end{align}
In this case, the three aligned conditions are $\cos\theta_{1+} =+1$, $\cos\theta_{2-}=+1$ and $\cos\theta_{2+}=-1$, corresponding to cases \ref{widesecondary}, \ref{coincidentsecondarysmall}, and \ref{updown} above. This is a BH binary where the secondary spin undergoes wide nutation, the morphology transitions from C$+$ to C$-$, and the up-down configuration is part of the precession cycle.

Two examples are shown in Fig.~\ref{doubletr}. For $q=0.95$, $\chi_1=0.3$, and $\chi_1=0.8$, triple transitions as predicted by Eq.~(\ref{triple1}) take place at $r=r_{\rm wide}\simeq 85$ if $\chi_{\rm eff}\simeq-0.24$ and $\kappa\simeq -0.11$. For $q=0.95$, $\chi_1=0.8$, and $\chi_1=0.5$, triple transitions as predicted by Eq.~(\ref{triple2}) take place at $r=r_{\rm wide}\simeq 42$ if $\chi_{\rm eff}\simeq0.17$ and $\kappa\simeq 0.09$.

\section{Conclusions}

Morphological transitions are generic relativistic effects in the dynamics of BH binaries. Single transitions were uncovered a decade ago~\cite{2015PhRvD..92f4016G}, with hints dating back almost two decades~\cite{2004PhRvD..70l4020S}. In this paper, we show that a  more restrictive set of sources might undergo two (or even three) concurrent morphological transformations. As shown in this contribution, it turns out that all such cases can be mapped and characterized analytically. We confirmed these analytical insights with numerical integrations using the {\sc precession} code; results from an ensemble of binaries are presented in Fig.~\ref{distrfour}, while some targeted evolutions are shown in~Fig.~\ref{doubletr}.

It is worth noting that double—and even more so triple—transitions are not generic, but only affect a small portion of the BH binary parameter space. That said, any ensemble of GW sources will include some binaries passing through these configurations. This point is illustrated in Fig.~\ref{distrfour}, which shows that the C$-$ morphology becomes accessible only at a finite separation during the binary inspiral. This special separation is nothing but $r_{\rm wide}$ from Eq.~(\ref{rwide}). As shown in this paper, this is the largest separation where genuine coincident transitions (thus excluding cases~\ref{updown} and~\ref{impossible}) can take place.
Somewhat heuristically, Ref.~\cite{2015PhRvD..92f4016G} found that multiple (though not concurrent) transitions—which we now understand result in binaries in the C$-$ morphology—are more likely for systems with roughly equal masses and unequal spin magnitudes. Indeed, these are precisely the conditions that maximize $r_{\rm wide}$, thus providing an analytical foundation for those empirical findings.

There have already been a few attempts at directly constraining spin morphologies from both current and simulated GW data~\cite{2022PhRvD.106b4019G,2023PhRvD.108j3003J}. At present, these have largely returned null results because spin-spin effects are highly subdominant, and higher signal-to-noise ratios are necessary. Encouragingly, some recent studies have been presented where multiple GW events are analyzed at the population level~\cite{2025arXiv250204278B,2022PhRvL.128c1101V}, though these have been framed in the somewhat outdated language of spin-orbit resonances~\cite{2004PhRvD..70l4020S,2013PhRvD..87j4028G} instead of the more modern concept of spin morphologies (the former are nothing but a specific limit of the latter).

As the field of observational GW astronomy progresses toward more numerous and louder events, it might become possible to infer the existence of the C$-$ morphology—which would indirectly point to the existence of the double transitions studied here—or, admittedly with a pinch of luck, even observe a BH binary undergoing such double transitions, once more uncovering the beautiful complexity of the two-body problem in General Relativity.

\section*{Acknowledgements}
D.G. and E.B. thank the City of Beijing, the Yanqi Lake Beijing Institute of Mathematical Sciences and Application (BIMSA), and the International Congress of Basic Sciences (ICBS) for selecting {\it ``Are merging black holes born from stellar collapse
or previous mergers?''}, Phys.~Rev.~D 95, 124046, (2017)~\cite{2017PhRvD..95l4046G} for a 2025 Frontiers of Science Award (FSA) in Physics. 

D.G., G.For., and G.Fum.~are supported by 
ERC Starting Grant No.~945155--GWmining, 
Cariplo Foundation Grant No.~2021-0555, 
MUR PRIN Grant No.~2022-Z9X4XS, 
Italian-French University (UIF/UFI) Grant No.~2025-C3-386,
MUR Grant ``Progetto Dipartimenti di Eccellenza 2023-2027'' (BiCoQ),
and the ICSC National Research Centre funded by NextGenerationEU.
D.G. is supported by MSCA Fellowship No. 101064542--StochRewind, 
MSCA Fellowship No.~101149270--ProtoBH, 
and MUR Young Researchers Grant No. SOE2024-0000125.
E.B.~is supported by NSF Grants No. AST-2307146, PHY-2513337, PHY-090003, and PHY-20043, by NASA Grant No. 21-ATP21-0010, by John Templeton Foundation Grant No. 62840, by the Simons Foundation, and by Italian Ministry of Foreign Affairs and International Cooperation Grant No.~PGR01167.
Computational work was performed at CINECA with allocations 
through INFN and Bicocca, and at NVIDIA with allocations through the Academic Grant program.

{\small

}

\vspace{15pt}

\address{[Gerosa, Foroni, Fumagalli] {Dipartimento di Fisica ``G. Occhialini'', Universit\'a degli Studi di Milano-Bicocca, Piazza della Scienza 3, 20126 Milano, Italy}}

\address{[Gerosa, Fumagalli]  { INFN, Sezione di Milano-Bicocca, Piazza della Scienza 3, 20126 Milano, Italy}}

\address{[Berti] { William H. Miller III Department of Physics and Astronomy, Johns Hopkins University, Baltimore, Maryland 21218, USA}}

\vspace{0.3cm}
\noindent E-mail address: 
\href{mailto:davide.gerosa@unimib.it}{davide.gerosa@unimib.it}

\end{document}